\documentstyle[12pt]{article}

\newcommand{\be}{\begin{equation}}
\newcommand{\ee}{\end{equation}}
\newcommand{\bea}{\begin{eqnarray}}
\newcommand{\eea}{\end{eqnarray}}

\newcommand{\SMI}{
                  \vspace{-10cm}
                  \begin{flushright}
                  gr-qc/9908047
                  \end{flushright}
                  \vspace{9.5cm}
                 }
\title{Entropy Bounds, Holographic Principle and Uncertainty Relation}
\author{
M. G. Ivanov\thanks{mgi@mi.ras.ru}\\
{\small \it Moscow Institute of Physics and Technology;
  Institutsky Per.9, Dolgoprudny, Moscow Reg., Russia} \\
{\small \it Department of Physics; University of California;
            94720-7300, Berkeley, CA , USA}
\\ and \\
I. V. Volovich\thanks{volovich@mi.ras.ru}\\
{\small \it Steklov Mathematical Institute;
 Gubkin St.8, 117966, Moscow, Russia}\\
}
\date{August 16, 1999}
\begin{document}
\maketitle
\SMI
\begin{abstract}
 A simple derivation of the bound on entropy is given
and the holographic principle is discussed.
 We estimate the number of quantum states inside space region on
the base of uncertainty relation.
 The result is compared with the Bekenstein formula for entropy bound,
which was initially derived from the generalized second law of
thermodynamics for black holes.
 The holographic principle states that the entropy inside a region
is bounded by the area of the boundary of that region.
 This principle can be called the kinematical holographic principle.
 We argue that it can be derived from the dynamical holographic
principle which states that
the dynamics of a system in a region
should be described by a system which lives on the boundary of the
region.
 This last principle can be valid in general relativity
 because the ADM hamiltonian reduces to the surface term.
\end{abstract}

\section{Introduction}

 There has been a great deal of interest recently in the Bekenstein
bound on entropy and the holographic principle as new and perhaps
fundamental principles in physics (see [1-20] and Refs. therein).
 According to Bekenstein \cite{b0} there exists a universal bound on the
entropy $S$ of any object of maximal radius $R$ and total energy $E$:

\be \label{bound}
	S\leq\frac{2\pi}{\hbar c}RE.
\ee

 The bound was derived from the requirement that
the generalized second law of thermodynamics for black holes
be respected when
a box containing entropy is placed without radial motion near the
horizon of Schwarzschild black hole and dropped into it \cite{b0}.
 Despite the derivation from gravitation gedanken experiment
the bound (\ref{bound}) does not involve the gravitational constant.

 The holographic principle \cite{hft,sssknd} states
that one has the following bound on the total entropy $S$
contained in a region of space bounded by the spatial
surface of the area $A$,
\be\label{A4}
	S\leq\frac{A}{4l_p^2},
\ee
where $l_p$ is the Planck length  $l_p=\sqrt{\hbar G/c^3}$.
 The bound (\ref{A4}) includes the gravitational constant $G$.

 There are already many discussions of bounds (\ref{bound})
and (\ref{A4}).
 However these important principles  deserve a further study.
 In this note the number of quantum states inside space region
is estimated on the base of uncertainty relation.
 The result is compared with the Bekenstein formula for entropy bound,
which was initially derived from the generalized second law of
thermodynamics for black holes.
 Then we discuss the holographic principle.
 The holographic principle states that the entropy in a region is
bounded by the area of the boundary of the region.
 This principle can be called the {\it kinematical}
 (or thermodynamical) holographic principle.
 We argue that it can be derived from the {\it dynamical} holographic
principle which states that
the dynamics of a system in a region
should be described by a system which lives on the boundary of the
region.
 Actually it is well known that the Hamiltonian in general relativity
may be reduced to the surface term.

 Another approach to holography based on
chaos is considered in \cite{amrv,hft2}.

\section{The bound on entropy}

 In this section we shall give a simple derivation
of the bound on entropy.
 Let us consider a region in 3-dimensional space of characteristic size $R$,
which contains energy $E$.
 We use ``natural'' system of units:
$$
  c=G=\hbar=1.
$$

 Due to the uncertainty relation the minimum energy $\varepsilon(R)$
of particle localized inside the region is of the order $1/R$,
\be
  \varepsilon(R)\sim 1/R,
\ee
(since $\varepsilon(R)\sim\sqrt{m_{min}^2+p_{min}^2}\sim p_{min}\sim 1/R$).
 The energy	$\varepsilon(R)$ can be considered as (minimum)
quantum of energy for region with radius $R$.

 The maximum number of particles inside the region for fixed $E$
could be estimated as maximum number of energy quanta, so
\be
	{\cal N}(E,R)\sim \frac{E}{\varepsilon(R)}\sim ER.
\ee

 Let us estimate the maximal entropy of the system.

 We have to count the number of quantum states corresponding
to given values of $E$ and $R$.
 If there is no degeneration of energy levels then our problem
is reduced to the counting of number of sets of positive
integers $(n_1,\dots,n_k)$ such that
\be
	\sum_{i=1}^k n_i\leq {\cal N}(E,R).
\ee
 Here $n_i$ is the number of energy quanta of the $i$-th particle.
 One can easily see that the number of such sets is
$2^{{\cal N}(E,R)}$.
Therefore we obtain for the entropy of a system the bound
$S\leq b{\cal N}(E,R)$, or, because ${\cal N}(E,R)\sim ER$,
\be\label{vv}
	S\leq bER,
\ee
where $b$ is a constant.
 So we have derived the
Bekenstein type bound (\ref{vv}) by using basically
only the uncertainty relation.

 The presence of finite number of internal degrees of freedom
and finite number of particle species, as well as
the degeneration of energy levels
due to the 3-dimensionality of space, does not change
the bound (\ref{vv})
(it can change only the constant $b$, see Appendix).

\section{Kinematical  holographic principle}
 If we assume that the size of the system $R$ with the energy (mass)
$E$ is greater than the Schwarzschild radius $2E$, i.e. $2E<R$, then
from the bound (\ref{vv}) one gets the bound
\be
	S\leq \frac{b}{2}R^2,
\ee
 which can be interpreted as the holographic principle \cite{hft,sssknd}.
 This principle says that the number of quantum degrees of freedom in a
region is bounded by the area of the boundary surface.
 In the such  form it can be called the kinematical holographic principle.

\section{The Bekenstein bound}

 The bound (\ref{vv}) is similar to the Bekenstein bound (\ref{bound}).
 However, this simple derivation does not fix
the constant $b$ in (\ref{vv}),
which depends on the particle spectrum.

 The factor $ER$ does not depend on dimensionality of space-time.
 Exact calculation in the case of thermal radiation gives the
factor $(ER)^{\frac{D-1}{D}}$.

 The maximum energy inside the region is of order $R$
($R^{D-3}$ in D-dimensional case)
\be
	E_{max}(R)\sim R.
\ee
For energies above $E_{max}$ the considered region is hidden under
horizon, and the consideration of the region from the point of view of
distant observer is senseless,
\be
	S_{max}(R)=\max_E S_{max}(E,R)=S_{max}(E_{max}(R),R)\leq b'R^2.
\ee

 Let us consider the spherical region of radius $R$, then $E_{max}=R/2$,
$S_{max}(R)=\frac{b}{2} R^2$.
 If we assume that entropy of black hole of
radius $R$ ($S_{b.h.}=\pi R^2$) is the maximum entropy of
the region of radius $R$, then it is natural to set
$b=2\pi$, and one gets
\be\label{2per}
	S_{max}(E,R)\leq 2\pi ER.
\ee

Formula (\ref{2per}) coincides with (\ref{bound}), nevertheless the
problem to calculate the proportionality coefficient $b$ requires
a special discussion.

 We have to distinguish between
derivation of entropy bound by black hole and
field theory arguments.
 The bound derived using black hole arguments represent the maximum entropy
of system which can be absorbed by black hole.

 For example, let us consider absorbtion of energy $E$ and entropy S
by Schwarzschild black hole of mass $M$.
 The change of black hole entropy is
\be
	\delta S_{b.h.}=4\pi(E^2+2ME)\geq S.
\ee
 Relation $\delta S_{b.h.}\geq S$ does not mean that entropy of any
system of energy $E$ has to be less than $\delta S_{b.h.}$, actually
$\delta S_{b.h.}$ is the maximal entropy of the system of energy $E$,
which can be absorbed by black hole of mass $M$.
 If $M$ is small enough $S$ can be greater than $\delta S_{b.h.}$,
but in this case absorbtion is impossible.

 The bound (\ref{bound}) is independent on mass of black hole,
which is supposed to be large in comparison with $E$, but even this
independence does not allow us to conclude that the bound could
not be interpreted in similar way.
 We could imagine that to force system to be absorbed by black
hole we have to increase its energy and/or decrease its entropy,
or that any box contained the system would be destroyed by
pressure of the system.
 The other possible way is to consider the bound as necessary
condition, which has to be valid for any theory compatible with
gravity.
 So, to postulate the bound (\ref{bound}) we have to appeal to
some extra arguments besides the gedanken experiments with
black holes.

\section{Interpretations of the Bekenstein bound}

	There is the very common implicit assumption: 
{\em any system can be
absorbed by black hole, if black hole is large enough}.
 Actually this assumption is not obvious.
 One can consider for example the gas uniformly distributed
in the space such that for its collapse or for the absorbtion
by the black hole one could require infinite time.

	Let us summarize possible interpretation of bound (\ref{bound}):\\
(i) $2\pi RE$ is the maximum possible entropy for
        any selfconsistent theory;\\
(ii) $2\pi RE$ is the maximum possible
       entropy for any theory with gravity;\\
(iii) $2\pi RE$ is the maximum possible
    entropy for any theory which admits
	  black hole solutions;\\
(iv) $2\pi RE$ is the maximum possible entropy of system, which can be
	absorbed by some black hole.

	Interpretations (i) and (ii) were mentioned
by Bekenstein in  \cite{b-b}.

 There is no direct observation of black hole, so the interpretation
(iii) also can not be skipped without discussion.

	Interpretation (iv) was not formulated explicitly, 
but spirit of this
interpretation is present in another Bekenstein paper \cite{b1}.
It was proved that $2\pi RE$ becomes maximal 
entropy of the system
of large number of particle species only
if we take into account energy of the box restricted the system,
i.e. system with entropy larger then $2\pi RE$
could be absorbed by black hole
only inside the box, which energy is large enough
to satisfy Bekenstein bound for whole system.
 This example demonstrates the fashion of trick,
which could be used by nature in the case of
interpretation (iv) to protect second law of thermodynamics.
 In special  relativity there is no sense a notion
of perfect solid body as well as a notion of a massless
box in quantum field theory.
 We do not know which abstractions will be obsolete in
quantum gravity.
 Maybe we have to take into account energy exchange
between the system in the box and space-time,
or maybe an infinitly high potential barrier
around internal space of the box is unacceptable
approximation.

 An alternative bound was also suggested by Unruh and Wald
\cite{uw1},
\be\label{uw}
	S\leq V s(E/V).
\ee
 According to the Bekenstein paper \cite{b1} the Unruh and Wald bound
(\ref{uw}) introduced in  \cite{uw1} is neither necessary,
nor sufficient.
 Nevertheless it is natural to assume, that thermal
radiation is maximally entropic in the classical case
(low energy, intermediate volume without gravity effects).
 It is the well known fact
(see \cite{nf}), that
for $V<V_{cr}$,
\be
	V_{cr}=const E^5,
\ee
thermal radiation becomes unstable and partially collapses into
black hole.
 For a given volume and energy, if $V>V_{cr}$ then
we can expect that the system with only thermal radiation
will have the maximum entropy.
 In this case the Unruh and Wald bound (\ref{uw}) is valid.
 However in the case $V<V_{cr}$, the system should have
the maximum entropy if it contains the
combination of thermal radiation and black hole.
Bekenstein bound probably covers both ranges of $V$.

\section{Dynamical holographic principle}

 One may formulate also the principle that the dynamics of a physical
system in a region which includes gravity should be described by a system
which lives on the boundary of the region.
 Such  principle can be called the dynamical holographic principle.
 It is discussed in the context of the AdS/CFT correspondence
\cite{mld,bhtz,sm,sw,ar,gdd}.

 We would like to point out that the dynamical holographic principle
in a certain sense is valid in classical general relativity.
 This is due to the well known fact that the density
of the ADM Hamiltonian $H(x)$ in general relativity is the total
divergence.
 One has (see e.g. \cite{nf,hh})
\be
	H(x)=T_0(x)-\partial_i\partial_k q^{ik},
\ee
where
\be
	T_0(x)=q^{ij}q^{kl}\left(\pi_{ik}\pi_{jl}-\pi_{lj}\pi_{kl}\right)
	+g_3 R_3
\ee
with
\bea
	q^{ik}=h^{i0}h^{k0}-h^{00}h^{ik},~~
	h^{\mu\nu}=\sqrt{-g}g^{\mu\nu}.
\eea
 By the constraint
\be
	T_0=0
\ee
 the Hamiltonian reduces to the surface term
\be
	H=\frac{1}{2}\int H(x) d^3x=
	-\frac{1}{2}\int\partial_i\partial_k q^{ik}d^3x
	 =-\frac{1}{2}\int\partial_k q^{ik}d\sigma_i.
\ee
 This means that all dynamical information is encoded in
the data on the boundary surface.
 Of course the system is highly nonlinear because we have to
solve constraints but in principle  one has the dynamical
holography in classical general relativity.

	One gets the similar conclusion if one uses the pseudo-tensor
\cite{ll} which leads to the expression for the energy-momentum
as the integral over the boundary surface
\be
	P^{\mu}=\int h^{\mu\nu\lambda}d\sigma_{\nu\lambda},
\ee
where
\be
	h^{\mu\nu\lambda}=\frac{1}{2}\partial_\sigma
 \left((-g)
     (g^{\mu\nu}g^{\lambda\sigma}-g^{\mu\lambda}g^{\nu\sigma})
 \right).
\ee

 Now by having the classical dynamical holographic principle we can
argue for the quantum kinematical principle.
 In fact, if the energy and other dynamical quantities of the
system are expressed as the integrals over the boundary surface
then the number of quantum states of the system should be
estimated by the area of the surface.
 We have to assume a cutoff
on the surface to remove divergences.

\section{Conclusion}

In this note the bounds on entropy and the holographic principle
have been discussed.
 By using the simple counting arguments and uncertainty relation
we have demonstrated that a general Bekenstein type bound (\ref{vv})
is valid.
 It is pointed out also that one has to distinguish between
the kinematical and dynamical holographic principle and that the
last one is actually valid in classical general relativity.
 We have argued that the quantum kinematical holographic principle can be
derived from the classical holography. However,
it is an open question whether this classical holography can be
used for the rigorous justification of the quantum holographic
principle because of the problem of divergences.

\section*{Appendix}
 It is convenient to use the notion of entropy as deficit of information
we need to describe the system state.

 Let us assume that we have to take
into account only finite number of particle species, each sort
of particle has finite number of internal states.
 This assumtion seems to be natural
for large $R$ (i.e. for small $\varepsilon$).
 So we need only finite number of bits of information to describe
internal degrees of freedom of one particle.
 The maximal number of particles is about ${\cal N}(E,R)$,
so we need about ${\cal N}(E,R)$ bits of information. 

Due to uncertainty principle we can determine momentum with accuracy
of $\varepsilon(R)$, so $x$ projection of momentum of the $i$-th
particle  can be described by natural $n_i^x$ and 
sign ``$+$'' or ``$-$''.
We can describe these signs using ${\cal N}(E,R)$ bits of information.
One has
\bea
	p_i^x&=&\pm n_i^x \varepsilon(R),\\
	E_i&\geq&|p_i^x|,\\
	E&=&{\cal N}(E,R)\varepsilon(R)=
	\sum_iE_i
	\geq\sum_i n_i^x\varepsilon(R)
\eea
so
\be\label{summa}
	{\cal N}(E,R)\geq\sum_i n_i^x.
\ee
To count the number of possible sets of $n_i^x$, which satisfy
relation (\ref{summa}) we  consider the following string
\be
	 (1w1w1w\dots w1w1w0)\cdot0,
\ee
which contain ${\cal N}(E,R)$ symbols ``1'' 
and ${\cal N}(E,R)$ symbols ``$w$''.
To describe any possible sum of the 
form (\ref{summa}) we have to replace all
``$w$'' by the following strings: ``$+$'' or ``$)+($''.
We can use just $3\cdot 2\cdot{\cal N}(E,R)$ bits to describe
momenta.

 Finally, for large $R$ we get
\be\label{bER}
	S_{max}(E,R)\leq b ER.
\ee
where the constant $b$ does not depend on $E$ and $R$.

\section*{Acknowledgements}

 We are grateful to I.Ya.Aref'eva and B.Dragovich
for useful discussions.
 I.V.V. is supported in part by INTAS grant 96-0698
and RFFI-99-01-00105.

\end{document}